**Sexiphenyl on Cu(100): nc-AFM tip functionalization and identification**

*Margareta Wagner\*, Martin Setvín, Michael Schmid, and Ulrike Diebold*

Institute of Applied Physics, TU Wien, Wiedner Hauptstrasse 8-10/134, 1040 Wien, Austria

\* Corresponding author: wagner@iap.tuwien.ac.at

*Abstract*

The contrast obtained in scanning tunneling microscopy (STM) and atomic force microscopy (AFM) images is determined by the tip termination and symmetry. Functionalizing the tip with a single metal atom, CO molecule or organic species has been shown to provide high spatial resolution and insights into tip-surface interactions. A topic where this concept is utilized is the adsorption of organic molecules at surfaces. With this work we aim to contribute to the growing database of organic molecules that allow assignment by intra-molecular imaging. We investigated the organic molecule *para*-sexiphenyl ($C_{36}H_{26}$, 6P) on Cu(100) using low-temperature STM and non-contact AFM with intra-molecular resolution. In the sub-monolayer regime we find a planar and flat adsorption with the 6P molecules rotated 10° off the <010> directions. In this configuration, four of the six phenyl rings occupy almost equivalent sites on the surface. The 6P molecules are further investigated with CO-functionalized tips, in comparison to a single-atom metal and 6P-terminated tip. We also show that the procedure of using adsorbed CO to characterize tips introduced by Hofmann et al. Phys. Rev. B 112 (2014) 066101 is useful when the tip is terminated with an organic molecule.

# 1. Introduction

Sexiphenyl ($C_{36}H_{26}$, Hexaphenyl, 6P) is a technologically important organic semiconductor, which is appreciated for its blue light emission in electroluminescence [1] and its photovoltaic properties [2]. The molecule has a rod-like shape consisting of six phenyl rings in *para*-configuration and Van der Waals dimensions of 2.72 × 0.67 $nm^2$ [3]. Having a well-defined interface between the 6P film and the metal back electrode is important for the performance of devices, e.g. the charge injection into the organic layer. Thus, the adsorption and growth of 6P has been investigated on many metal surfaces including several low-index surfaces of fcc metals such as Cu [4], Ag [5-7], Au [8,9] and Al [10,11]. In the sub-monolayer range the adsorption configuration is determined by the interaction between the molecules and the substrate. This can lead to different molecular configurations even on structurally similar surfaces, for example on the anisotropic (110) surfaces of Cu [4] and Ag [6, 7] (unit cell dimensions of Cu: 0.2556×0.3601 $nm^2$ and Ag: 0.2889×0.4086 $nm^2$). In both cases 6P adsorbs planar and flat. On Cu(110) the molecule aligns its long axis parallel to the densely packed atomic rows [4], but on Ag(110) 6P is oriented perpendicular to these rows. On the basis of band structure measurements this behavior was explained as due to differences in the surface state and thus a different hybridization of the lowest unoccupied molecular orbital [7]. On fcc(111), molecular alignment along the high-symmetry directions was observed on Al [10,11], on Cu [7] and Ag [5,7], in either planar or twisted configurations. On Au(111), the reconstruction is not lifted by the first layer 6P. Deposition at RT led to molecules rotated 3° off of the <11-2> (i.e., 30° off of the close-packed directions) surface directions. Increasing the substrate temperature to 330 K during deposition resulted in domains with <1-10> and <11-2> orientation, and only <1-10> orientation was observed [8] at 430 K. Interestingly, the lattice parameters of three of these (111) surfaces are very similar, with Cu deviating from it (Al: 0.286 nm, Ag: 0.2889 nm, Au: (unreconstructed) 0.2884 nm, Cu: 0.2556 nm). In general, it should not be too surprising if 6P on Cu(100) is aligned in a different fashion from what has been observed on other fcc(100) surfaces. On Ag(100), for example, measurements of the electronic band structure and momentum maps suggest 6P in two different configurations, adopting either the close-packed <011> and <010> directions or those rotated by 22.5° with respect to these directions [7]. The interatomic distance on Cu(100) measures 0.2556 nm, in contrast to 0.2889 nm for Ag(100). A <011> oriented 6P molecule thus covers 11 and 10 substrate atoms, respectively.

Non-contact atomic force microscopy (nc-AFM) provides new and in combination with STM complementary insights into the physics of surfaces and adsorbed species. Functionalizing the tip with a CO molecule has been shown to provide high lateral resolution on organic species when probing short-range forces. In repulsion, the carbon backbone of the molecule is imaged, including the "bonds" [12]. The mechanism behind this contrast is based on the flexible attachment of the CO molecule to the tip, so it can bend away from repulsive maxima in the surface potential [13]. Moreover, in all scanning probe techniques the roles of "tip" and "surface" are reversed when a small probe species on the surface is scanned by a blunt tip. By using a CO molecule for this purpose the symmetry of the tip can be determined in nc-AFM, which is known as "CO front atom identification", COFI [14-16].

In this work the adsorption of 6P in sub-monolayer coverages on Cu(100) is investigated with scanning tunneling microscopy (STM) and non-contact atomic force microscopy (nc-AFM) with differently terminated tips. High resolution on the molecular structure was obtained with CO-functionalized tips, and we extend the COFI from atom identification to "CO front molecule identification".

## 2. Experimental

The experiments were performed in a two-chamber UHV system equipped with an LT-STM/AFM (ScientaOmicron) operating at 5 K, as well as standard sample cleaning and preparation tools. The nc-AFM were conducted with qPlus tuning-fork sensors [17] with an additional wire for the tunneling current [18] and a differential preamplifier, described in ref. [19]. Two different sensors were used: (1) $f_R$ = 30,000 Hz, k ≈ 1,800 N/m, Q ≈ 2,200, and (2) $f_R$ = 47,500 Hz, k ≈ 3,750 N/m, Q ≈ 10,000. The oscillation amplitude was <100 pm in the constant-height measurements. The tips, made from electrochemically etched W wire, were glued to the tuning fork. They were initially prepared by self-sputtering, and then by gentle voltage pulses during the experiments. In the constant-height AFM measurements, the tunneling current was simultaneously acquired by applying a small bias voltage. The Cu single crystal was cleaned by cycles of sputtering (Ar$^+$ ions, 1 keV) and annealing in UHV at 500 °C, resulting in atomically flat and clean surfaces. Sexiphenyl was deposited at room temperature from a homebuilt evaporator via thermal sublimation from powder (Tokyo Chemical Industry Co., Ltd.) at ~230 °C (thermo-couple reading). The deposition rate was monitored with a quartz crystal microbalance. After deposition, the sample was transferred into the LT-STM. Small amounts of CO were dosed directly into the STM/AFM at 5-6 K by backfilling the chamber and opening a window in the radiation shields of the cryostat.

## 3. Results and Discussion

An overview of the sub-monolayer coverage of 6P on Cu(100) is given in Figure 1(a). The 6P molecules are oriented in four directions and their appearance suggests a flat-lying adsorption geometry. In order to relate the molecular orientations to the Cu(100) substrate directions, the uncovered surface was imaged with atomic resolution in constant-height nc-AFM, see Figure 1(b). This makes apparent that the 6P molecules are oriented ±35° with respect to the [011] and [01-1] surface directions, i.e., their molecular axes include ±10° with the <010> type directions. The directions of the substrate and the 6P molecules are both indicated in Figure 1(b). The 6P orientation on Cu(100) is − as is the case for the (110) surface as well − different from what has been found on Ag(100). There, two phases were observed with the 6P molecules adopting either the <011> and <010> type orientations, or rotated by 22.5° with respect to these directions [7].

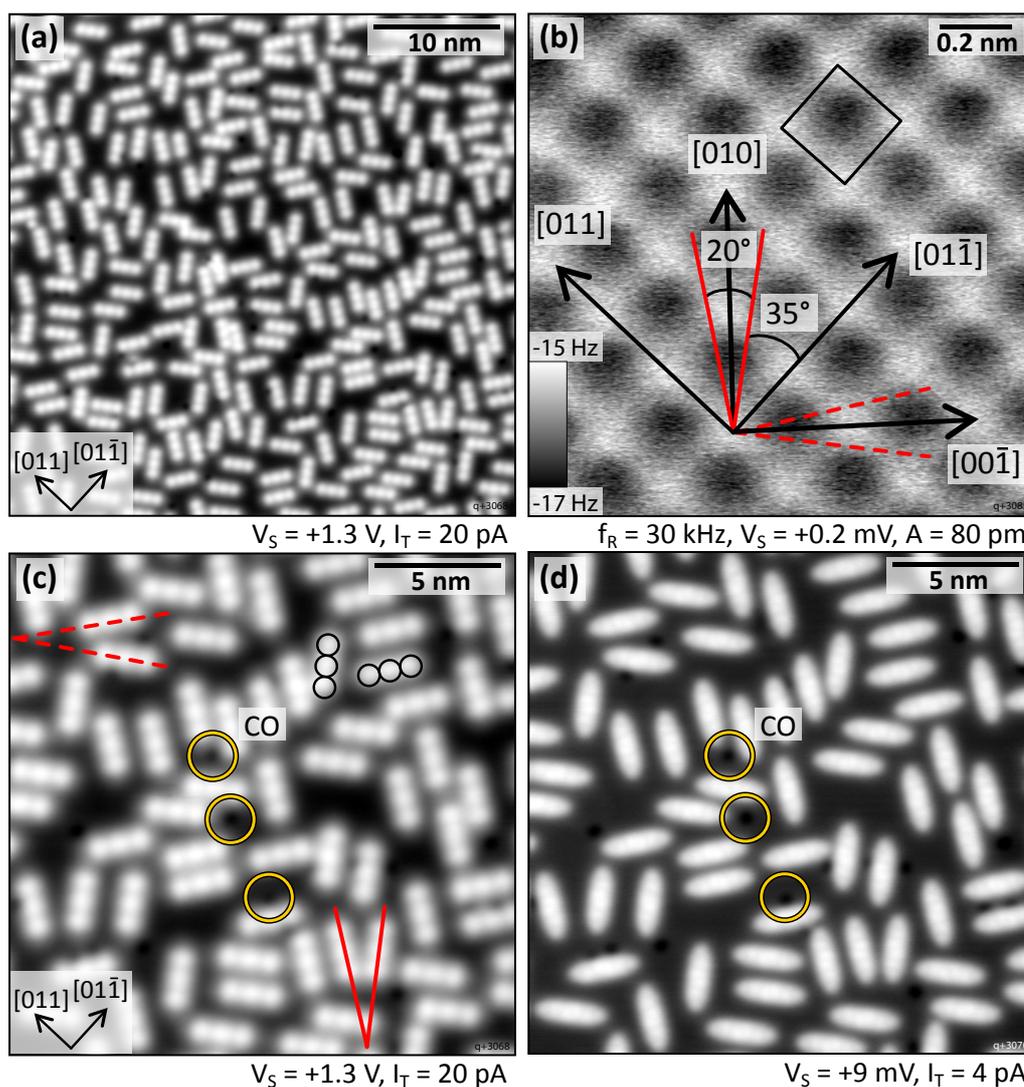

Figure 1: STM image of 6P/Cu(100), adsorbed at RT. (a) Overview. (b) Atomic resolution on the Cu surface in nc-AFM, the square unit cell is indicated. The substrate directions are labeled (black) together with the orientations of the 6P molecules (red). (c) Detailed view, constant-current STM. (d) Quasi-constant height STM of the same region as in (c). Several CO molecules are visible as small dark dots (yellow circles). All images were taken at 5 K.

The adsorption site of the 6P molecules is evaluated in Figure 2 by extracting the Cu(100) lattice from quasi-constant-height STM measurements by a 2D Fourier transform. In Figure 2(a) the Cu lattice is marked by dots, although it remains unknown whether the dots mark the top or hollow sites. The inset of Figure 2(a) shows an enlarged view of one molecule with the Cu lattice sites along the molecule highlighted by large red and white dots. For a planar and flat 6P molecule, four of the phenyl rings adopt very similar sites on the Cu lattice (red; phenyl rings 1, 3, 4 and 6, either top or hollow), while the remaining two phenyl rings are located in bridge sites (between the white dots; phenyl rings 2 and 5). Thus we conclude that the molecules adsorb in a configuration where the sites occupied by the majority of the phenyl rings (red dots) is favored. A schematic of 6P on Cu(100) for the two possible adsorption sites is given in Figure 2(b).

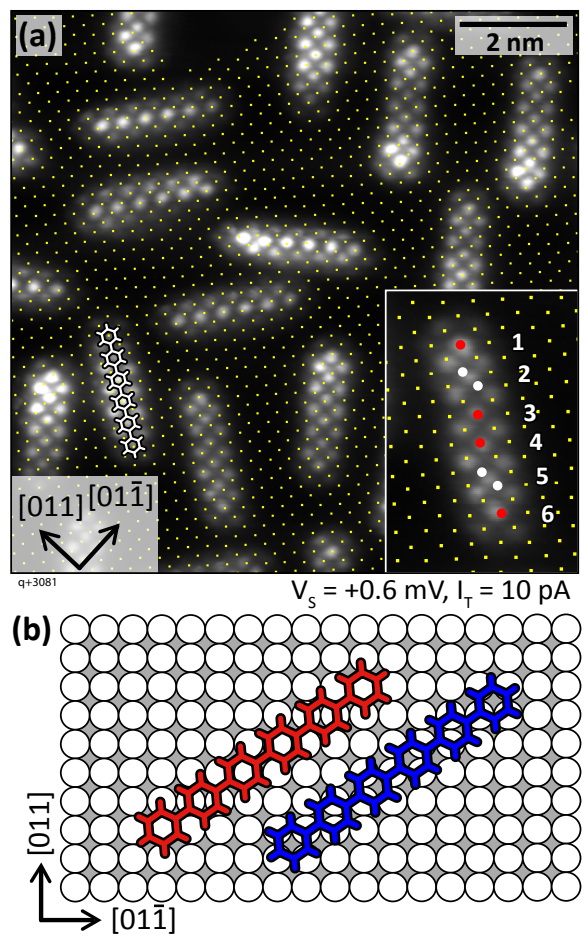

Figure 2: Adsorption site of 6P on Cu(100) (a) Quasi constant-height STM image. The Cu lattice extracted from the image by FFT is superimposed by yellow dots. The dots mark either the top or hollow position of Cu(100). The inset shows one of the molecules where the lattice dots along the molecule are highlighted in red (for the phenyl ring 1, 3, 4 and 6, top or hollow site) and white (rings 2 and 5, bridge site). (b) Atomic model of Cu(100) with 6P superimposed in the two possible configurations: on-top (red) or hollow sites (blue).

In the close-up STM image of the molecules in Figure 1(c), each 6P molecule features three protrusions (indicated by small black circles) along their long axis. Such an appearance is usually observed for 6P with alternatingly tilted phenyl rings, i.e., a twisted molecule. In such a configuration the three pairs of phenyl rings, i.e., the three repeating units along the molecule that are equally tilted can be identified in STM [4,20]. On Cu(100), however, as is shown below, 6P adsorbs planar and flat, and the three blobs are an electronic feature either induced by the tip or the adsorption site. As discussed earlier, the configuration of 6P on Cu(100) is such that the two phenyl rings in the center and those at the ends of the molecule have almost the same registry (Figure 2). Thus, also the interaction and overlap of the π-system with the substrate are expected to be similar, which could give rise to the three protrusions frequently observed in the constant-current STM images. The distance between the nodes is ~0.75 nm in the STM images. The same area as shown in Figure 1(c) was also imaged in quasi-constant height STM at a much lower bias voltage, see panel (d), where the molecules appear as oval objects with very low contrast of the intramolecular features.

CO molecules, dosed into the STM head at low temperatures, are visible as dark dots in Figures 1(c, d). They can be picked up with the AFM tip by manually approaching the tip to the surface while scanning in constant-height AFM mode. By applying a small bias voltage of a few mV, the mean tunneling current $<I_T>$ is acquired simultaneously to the AFM signal. A successful tip termination with CO is accompanied by a drop in tunneling current and a change in the AFM contrast due to the now extended, but weakly conducting CO tip. Figure 3 displays images of 6P on Cu(100) acquired with CO-functionalized tips. Here, the molecular structure appears bright due to the repulsion between the molecule and the tip. The tunneling current measured on top of the molecules was <0.02 nA (bias voltage +4 mV, repulsive regime) in panel (b), where the tip was closest to the surface. The six hexagons representing the phenyl rings are clearly visible, strongly suggesting a planar and flat adsorption geometry of the molecule. Note that the phenyl rings in the middle of the molecule appear distorted compared to those at the ends. This artifact is commonly observed in the nc-AFM imaging with CO tips. It varies for different substrate materials, and has been assigned to the lateral forces bending the flexible CO molecule at the tip [21,13].

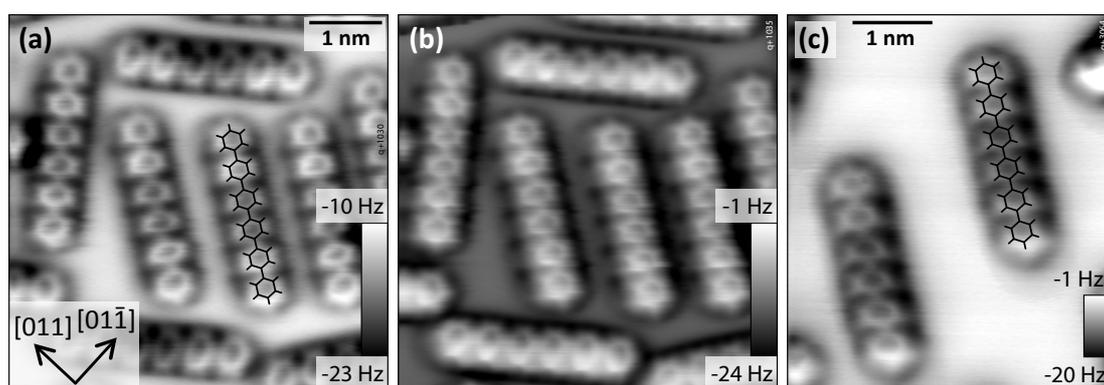

$f_R$ = 47.5 kHz: $V_S$ = +10 mV, A = 96 pm; $V_S$ = +4 mV, A = 45 pm     $f_R$ = 30 kHz, $V_S$ = +400 pV, A = 70 pm

Figure 3: Constant-height nc-AFM images of 6P on Cu(100) imaged with a CO-terminated tip. (a) and (b) shows the same surface area with different oscillation amplitude, leading to a lower tip-sample distance in (b). (c) A high-resolution image. The 6P structure is shown as overlay on one molecule in (a) and (c).

In comparison to the constant-height images obtained with the CO tip on 6P/Cu(100), Figure 4(a) shows images acquired with a metallic tip. The molecules are imaged dark due to attractive forces, with six, slightly brighter features along their axes. Note that one of the molecules in the lower center of the image consists of only five phenyl rings. The signal of the mean tunneling current $<I_T>$ is displayed in Figure 4 (b), and reveals tunneling currents of ~0.8 nA on top of the molecules. The CO molecules in-between the 6P molecules appear as spherical, bright dots in the AFM image of Figure 4(a), but are hardly visible in the current-image, panel (b). The inset in (b) shows the labeled/encircled CO from panel (a) with enhanced contrast. The CO molecules appear as "sombreros" in the (quasi) constant-height STM images, i.e., a bright protrusion surrounded by a dark boundary, see Figure 1(d) and Figure 4(b), and as dark dots in the constant-current STM image of Figure 1(c).

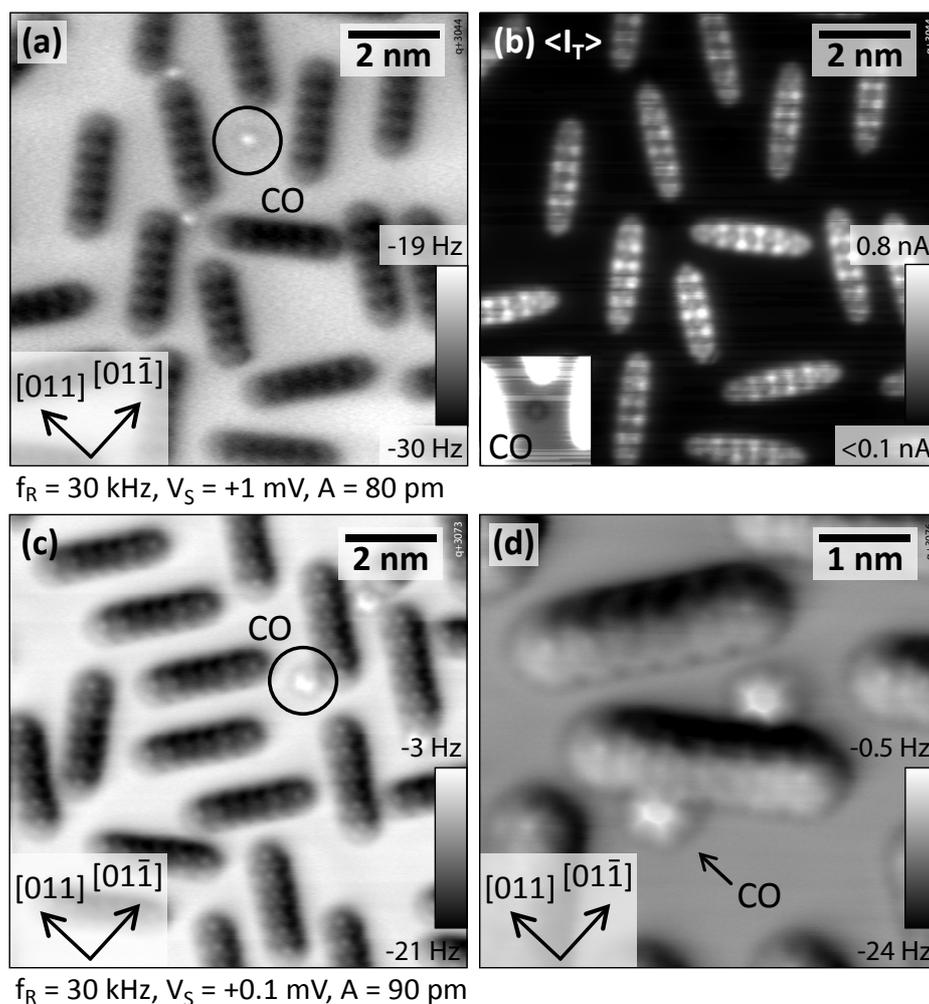

Figure 4: nc-AFM images of 6P and CO with different tip terminations. (a) Unknown, metallic tip. The CO molecules are imaged as small, bright dots. (b) Tunneling current signal recorded simultaneously with (a). (c, d) The CO molecules adsorbed on the surface reveal the tip termination as a benzene ring is adsorbed on the tip apex.

Functionalizing the tip by picking up a CO molecule is not always successful, especially in the presence of large molecules, which can be pushed around and picked up by the tip. Figure 4(c, d) shows the result of an attempt to pick up a CO molecule, where a 6P molecule (or a fragment) was picked up instead, resulting in a benzene ring attached to the tip apex in flat geometry. In the images obtained with this tip, the 6P can be still recognized, featuring six repeating units along its long axis. However, the small CO molecules co-adsorbed on the Cu(100) surface reveal the true nature of the tip in the CO front atom identification (COFI) measurement [14-16]. Acting now as the sharp "tip", the CO adsorbed on the surface scans the tip apex and thus the benzene ring is visible on the surface where the CO are adsorbed.

## 4. Summary

In this work we presented the adsorption of the organic molecule sexiphenyl on Cu(100) at room temperature, investigated at low temperatures with STM and nc-AFM. The tip was functionalized with a CO molecule and compared to a metal atom and (fragments of) a 6P molecule attached to the tip. The symmetry of the tip was identified with the COFI method, i.e., by imaging CO molecules on the surface. At sub-monolayer coverages we find that the 6P molecules align ~10° off of the high symmetry <011>-type directions in a planar and flat configuration. Although our analysis cannot distinguish between on-top and hollow sites, we can conclude that in this particular orientation the molecule has optimized its interaction with the substrate by having four of the six phenyl rings in a very similar adsorption site.

**Acknowledgements**

The following projects by the Austrian Science Fund (FWF) are gratefully acknowledged: T759-N27 (M.W.), and Z250 (U.D.).